\documentclass[useAMS,usenatbib,usegraphicx]{mn2e}

\newcommand{\chandra}{{\it Chandra}}
\newcommand{\asca}{{\it Asca}}
\newcommand{\xmm}{{\it XMM-Newton}}
\newcommand{\epic}{{\it EPIC}}
\newcommand{\pn}{{\it pn}}
\newcommand{\mos}{{\it MOS}}
\newcommand{\eso}{{\it ESO 323-G77}}
\newcommand{\chiqua}{$\chi^2_{\nu}$} 
\newcommand{\cgs}{erg~cm$^{-2}$s$^{-1}$}

\title[The complex X-ray spectrum of ESO~323-G077]{Detection of blueshifted emission and absorption and a relativistic Iron line in the X-ray spectrum of ESO~323-G077 \thanks{Partially based on observations obtained with XMM-Newton, an ESA science mission with instruments and contributions directly funded by ESA Member States and NASA.}}
\author[E. Jim\'enez-Bail\'on   et al.]{E. Jim\'enez-Bail\'on$^{1,2}$\thanks{E-mail: elena@astroscu.unam.mx},    Y. Krongold$^{1}$, S. Bianchi$^{3}$, G. Matt$^{3}$,  M. Santos-Lle\'o$^{4}$, \and  
 E. Piconcelli$^{5}$,  N. Schartel$^{4}$\\
$^{1}$ Instituto de Astronom\'{\i}a, Universidad Nacional Aut\'onoma de M\'exico, Apartado Postal 70-264, 04510 Mexico DF, Mexico \\
$^{2}$ LAEFF-INTA, Apartado 50727, 28080 Madrid, Spain\\
$^{3}$ Universita Roma Tre, Via della Vasca Navale 64, I-00146, Roma, Italy\\
$^{4}$ XMM-Newton Science Operation Centre, ESAC, ESA, Apartado. 50727, 28080 Madrid, Spain\\
$^{5}$ Osservatorio Astronomico di Roma (INAF), Via Frascati, 33, 00040, Monteporzio Catone, Italy }
\begin{document}

\date{Accepted 1988 December 15. Received 1988 December 14; in original form 1988 October 11}


\maketitle

\begin{abstract}
We  report  on   the  X-ray  observation  of  the   Seyfert  1  galaxy
ESO~323-G077 performed  with \xmm.  The \epic\ spectra  show a complex
spectrum  with  conspicuous  absorption  and emission  features.   The
continuum emission can  be modelled with a power law  with an index of
$1.99\pm0.02$ in  the whole  \xmm\ energy band,  marginally consistent
with typical  values of Type-I objects.  An  absorption component with
an      uncommonly      high      equivalent      Hydrogen      column
(n$_H=5.82^{+0.12}_{-0.11}\times10^{22}$~cm$^{-2}$)  is  affecting the
soft  part  of  the   spectrum.   Additionally,  two  warm  absorption
components are also  present in the spectrum.  The  lower ionised one,
mainly imprinting  the soft  band of the  spectrum, has  an ionisation
parameter of $\log{U}=2.14^{+0.06}_{-0.07}$ and an outflowing velocity
of  v=$3200^{+600}_{-200}$~km/s.   Two  absorption  lines  located  at
$\sim$6.7 and  $\sim$7.0~keV can be  modelled with the  highly ionised
absorber. The ionisation parameter  and outflowing velocity of the gas
measured         are         $\log{U}=3.26^{+0.19}_{-0.15}$        and
v=$1700^{+600}_{-400}$~km/s,  respectively.   Four  emission lines
  were  also  detected in  the  soft  energy  band.  The  most  likely
  explanation  for these emission  lines is  that they  are associated
  with an outflowing gas with  a velocity of $\sim2000$~km/s. The data
suggest that the  same gas which is causing  the absorption could also
being   responsible  of   these  emission   features.    Finally,  the
\xmm\ spectrum shows the presence of a relativistic iron emission line
likely originated in  the accretion disc of a Kerr  Black Hole with an
inclination of  $\sim$25 degrees.  We  propose a model to  explain the
observed X-ray  properties which invokes  the presence of  a two-phase
outflow  with cone-like  structure  and  a velocity  of  the order  of
2,000-4,000~km/s. The inner  layer of the cone would  be less ionised,
or even  neutral, than the outer  layer. The inclination  angle of the
source would be lower than the opening angle of the outflowing cone.
\end{abstract}


\section{Introduction}

Fast winds or outflows of ionised gas are known to be common in active
galactic nuclei (AGN).  They have  been detected for the first time in
optical/UV   spectra.    In  the   X-ray   band,   \asca,  \xmm\   and
\chandra\ systematic studies of AGN  established that at least half of
the  active galaxies  host warm  absorbers (Piconcelli  et  al.  2005;
George et  al. 1998; Reynolds \&  Fabian, 1995; Blustin  et al.  2005,
Mckernan, Yaqoob \& Reynolds, 2007). The evidence of transversal flows
(Mathur et al.  1994; Crenshaw  et al.  2003, Arav 2004) indicate that
they are ubiquitous in AGN,  becoming detectable only in certain lines
of  sight.    Typically,  the  ionised   gas  has  a   temperature  of
$\sim$10$^6$~K,    a    density     ranging    from    $10^{22}$    to
$10^{23}$~cm$^{-2}$, and frequently outflowing with typical velocities
of several  thousand km/s (Kaspi et  al 2002; Krongold  et al.  2003).
However, highly  ionised absorbers are not so  common.  Recent studies
(e.g. NGC~1365 Risaliti et al.  2005; PG1402+261 Reeves et al.  2004;
NGC4051, Krongold  et al.  2007;  NGC~985 Krongold et al.  2008), show
evidence  of extreme  characteristics  of these  hot  absorbers, as  a
relativistic velocity of the  outflowing gas or a dramatic variability
of the absorbing  features, indicating that the gas  should be located
very close  to the nucleus of  the AGN. Highly  ionised absorbers have
been probed thanks  to the high sensitivity of  \xmm\ and \chandra\ in
the  2-10~keV. In  a  recent  review, Cappi  (2006)  reports from  the
literature a  dozen of  AGN with blue-shifted  Fe absorption  lines in
their X-ray spectra.  However, there  are only few cases in which more
than one absorption line associated to the same outflowing material is
measured.   Deep  studies of  this  kind  of  objects are  crucial  to
establish the presence of these highly ionised absorbers.

According to  the paradigm  of the  origin of AGN, the  large energy
output  of  these  systems   is  explained  by  radiatively  efficient
accretion onto a supermassive Black Hole.  This scenario predicts that
the broad relativistic  Fe emission line would be  a relatively common
feature in  the X-ray spectrum of  AGN.  The most  outstanding case is
MCG-6-30-15 (Tanaka  et al.  1995; Wilms  et al.  2002,  Fabian et al.
2002).  However, only  about 30\% of the objects  with proved evidence
of the  relativistic iron line have  been reported (see  Nandra et al.
2006 for a  recent review, Jim\'enez-Bail\'on et al.   2005; Miller et
al.   2007).  In  many cases,  the data  analysis suggest  that highly
ionised  and   highly  absorbed   material  can  mimic   the  spectral
characteristic   of   the  relativistically   iron   line  (Reeves   et
al.  2004). Both,  systematic of  complete  sample studies,  as well  as
accurate determination of the parameters of the broad iron lines based
on deep single  target analysis, are crucial to  understand the physics
of the accreting process (Guainazzi et al.  2006, Nandra et al. 2006).

ESO~323-G077  is  a nearby  (z=0.015),  bright  (V=13.6  mag), type  I
Seyfert  galaxy,  discovered  by Fairall  (1992). Winkler  (1992) also
  classified as Seyfert 1  galaxy, although more recently, Veron-Cetty et
  Veron  2006  classified it  as  a  1.2  Seyfert galaxy.  In  X-rays,
ESO~323-G077 was detected for the first time in the {\it Rosat All-Sky
  Survey}. In the hard band, the galaxy was detected in the {\it RXTE}
Slew Survey  (Revnivtsev et al.   2004).  The measured  luminosity was
L$_{3-20\,keV}\sim7\times10^{42}$~erg/s.  The  galaxy is also included
in  the  First  Integral  Catalog  (Beckmann et  al.   2006),  with  a
luminosity  of  L$_{2-100\,keV}\sim1.6\times10^{43}$~erg/s.   In  this
paper we present the results  on the analysis of the \xmm\ observation
of  the Seyfert  galaxy  ESO~323-G077, which  shows  evidence of  both
relativistic iron line and the  presence of highly ionised gas in both
the X-ray soft and hard band of its spectrum.

\section{Observations and Data Analysis}\label{sect:data}

The \xmm\ (Jansen et al.  2001) observation of \eso\ was performed the
7$^{th}$ of February, 2006  (obsid. 0300240501). The {\it thin} filter
and  the {\it  large window}  mode were  selected for  the  {\it EPIC}
exposures.   All \xmm\  data  were processed  with  the standard  {\it
  Science  Analysis System}, SAS,  v7.0.0 (Gabriel  et al.   2004) and
using the  most updated calibration  files available in  January 2007.
The  \epic\  event lists  were  filtered  to  ignore periods  of  high
background  flaring following  the  method proposed  by Piconcelli  et
al. (2004).   The exposures after background filtering  are 24.0, 28.2
and  28.2  ks for  \pn,  {\it  MOS1},  and {\it  MOS2},  respectively.
According to  {\it SAS}  task {\it epatplot},  no sign of  pile-up was
detected in any of the \epic\  observations. The {\it RGS} data do not
have enough signal-to-noise to perform spectral analysis and therefore
they were not considered in the analysis.

The  spectra were extracted  from circular  regions of  35\arcsec\ and
centred on the maximum emission  of the source. The background regions
were  extracted from circular  regions of  1\arcmin\ and  1\farcm7 for
\pn\ and {\it MOS}  observations, respectively. The background regions
were  selected to  be located  close  to the  source and  free of  any
contamination  source. After verifying  the {\it  MOS1} and  {\it MOS2}
agree  one with each  other, both  spectra were  combined to  obtain a
higher  signal--to--noise ratio.  After  background subtraction,  both
spectra, i.e.  the \pn\  and the {\it  MOS1-2} combined,  were grouped
such  that  each bin  contains  at  last 50  counts  per  bin. We  are
therefore able to use the  modified $\chi^2$ technique (Kendall et al.
1973)  in the  spectral analysis.   We  assumed a  flat $\Lambda  CDM$
cosmology with  ($\Omega_M,\Omega_\Lambda$)=(0.3,0.7) and a  value for
the Hubble constant of 70 kms$^{-1}$ (Bennett et al. 2003).

\subsection{Spectral Analysis}\label{sect:analysis}

We  have  performed the  spectral  analysis  of  the data  using  {\it
  Xspec}~v.12.3   (Arnaud   1996).    The   \pn\  and   the   combined
\mos\   spectra   have   been   fitted  simultaneously,   leaving   to
independently   vary  only   the  normalisations   of   the  different
components.   Otherwise indicated,  all  parameters are  given in  the
\eso\ rest  frame. The errors quoted  in this paper refer  to the 90\%
confidence  level  (i.e.  $\Delta\chi^2$=2.71;  Avni  1976).  For  the
spectral fitting of  the \xmm\ data we have  considered the absorption
due to  the Galaxy, fixing  the equivalent Hydrogen column  density to
7.4$\times10^{20}$~cm$^{-2}$ (Dickey  \& Lockman, 1990).

\begin{figure}
\includegraphics[width=60mm,angle=-90]{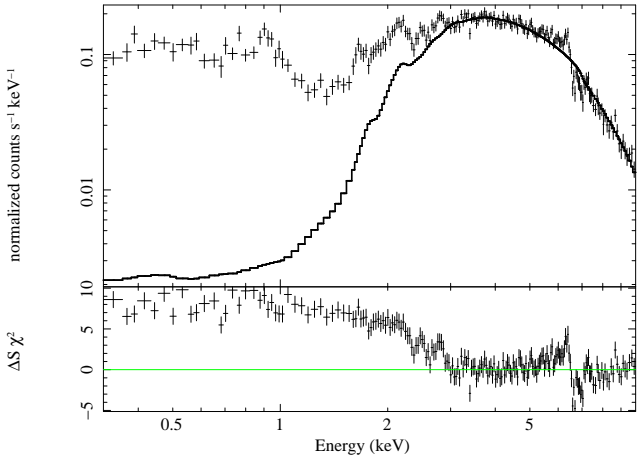}
\caption{\epic-\pn\  observed spectrum  and residuals  to  an absorbed
power law  model fitted  in the  3--10 keV and  ignoring the  5--8 keV
band.}\label{residuals_fe}
\end{figure}

Considering the  whole energy band (0.3-10~keV), a  power law modified
by  an  intrinsic  cold  absorption  component  does  not  provide  an
acceptable fit  to the  \epic\ data (\chiqua$\sim10$).   Both positive
and  negative residuals are  clearly present  below 3  keV and  in the
5--8~keV observed  energy band.  We therefore ignored  the data within
these bands and fitted a power law modified with intrinsic absorption.
We  measured  an  absorbing  equivalent  Hydrogen  column  density  of
$\sim6\times10^{20}$~cm$^{-2}$.  The  index of the  power law resulted
to be 1.88$\pm0.05$,  in good agreement with typical  values of type I
AGN (Piconcelli et  al.  2005).  Figure~\ref{residuals_fe} shows these
strong residual after re-notifying the ignored channels.  For the sake
of clarity, only \pn\ data are shown in the plot.

\subsubsection{The soft energy band}

In order to fit the conspicuous {\it soft excess}, we reintroduced the
energy bins below 3~keV.  We then added to the previous model an extra
power law  emission with the same  index of the one  associated to the
high energy band,  as expected for a  partial covering scenario.
The   additional   component  largely   improves   the  previous   fit
(\chiqua$\sim10$) with a resulting  \chiqua\ of 1.5. However, this new
fit  still  leaves  residuals   both  in  {\it  absorption}  and  {\it
  emission}.  We then tested the  presence of an extra cold absorption
component,  located  further  away  from  the  nucleus  and  therefore
affecting both components (i.e.  {\it wabs*(powerlaw+wabs*powerlaw)}).
However, an  ionised absorption  produced a more  adequate fit  to the
data than the cold one.  The warm absorption component has been fitted
using the modelisation  developed by Krongold et al.   2003, i.e.  the
{\it Phase} model.   The power law index of  the resulting model (i.e.
{\it   phase*(powerlaw+wabs*powerlaw)}   labeled   as   model   A   in
Tables~\ref{tab_mod1}          and          \ref{tab_mod2})         is
$\Gamma=1.90^{+0.07}_{-0.06}$   and  the   cold   absorption  Hydrogen
equivalent  column $6.34^{+0.18}_{-0.19}\times10^{22}$~cm$^{-2}$. 
  This value  is largely higher  than typical values of  Type-I active
  galaxies, which are mainly  compatible with absorption caused by the
  host  galaxy, i.e. $<10^{21}$~cm$^{-2}$  (Piconcelli et  al. 2005).
The  warm absorber    imprinting the  soft energy  band  has a
moderately        high        Hydrogen       equivalent        column,
n$_H$=$1.3^{+1.8}_{-0.8}\times10^{23}$~cm$^{-2}$   and  an  ionization
parameter of $\log{U}$ =$3.16^{+0.19}_{-0.12}$, where $U=\frac{Q}{4\pi
  R^2cn_e}$, Q  is the  luminosity of the  ionising photons, R  is the
distance to the source and $n_e$ the electron density.


\begin{figure}
\includegraphics[width=60mm,angle=-90]{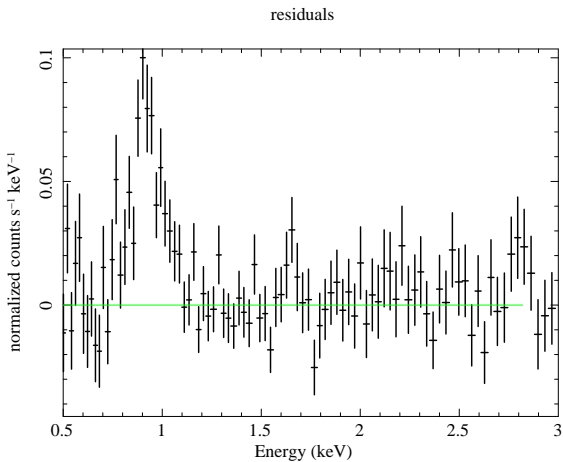}
\caption{\epic-\pn\  residuals to model A in the 0.5-3~keV
band. Emission features are visible at 1~keV,
1.7~keV  and 2.8~keV.}\label{line_res}
\end{figure}

However,  further positive  residuals are  also present  around 1~keV,
1.7~keV and  2.8~keV (see  Fig.~\ref{line_res}).  We have  measured two
emission   lines  around   $\sim$1~keV,  (0.93   and   1.04~keV,  i.e.
13.3~\AA\ and  11.9 \AA, respectively) which could   be naturally associated
with   the   Ne\textsc{ix}  triplet   (13.47;   13.55,  13.7\AA)   and
Ne\textsc{x}  at 12.13\AA\  with an  outflow velocity  of the  order of
2,000~km/s.  After the inclusion of  each of these emission lines, the
fit improved at a significant  level higher than 99.99\%, according to
the F-test.  Two more emission  lines, less significant but still with
probabilities of 95.9\% and 99.0\% are also detected, only in the {\it
  pn}  spectrum.  The  measured centroids  of the  lines  are 1.68~keV
(7.38\AA)   and  2.84~keV(4.37\AA)  and   could  be   associated  with
Mg\textsc{xi}  (7.85\AA) and  S\textsc{xvi} (4.73  \AA), respectively.
The outflowing  velocity of the  originating material would be  of the
order  of  20,000~km/s.   However,  an  outflow  with  a  velocity  of
$\sim2000$~km/s would  also be possible  if lines are  identified with
other species  (see Discussion  for further details).  The interesting
parameters of the lines are given in Table~\ref{tab_emlines}. 

\subsubsection{The hard energy band}

\begin{figure}
\includegraphics[width=60mm,angle=-90]{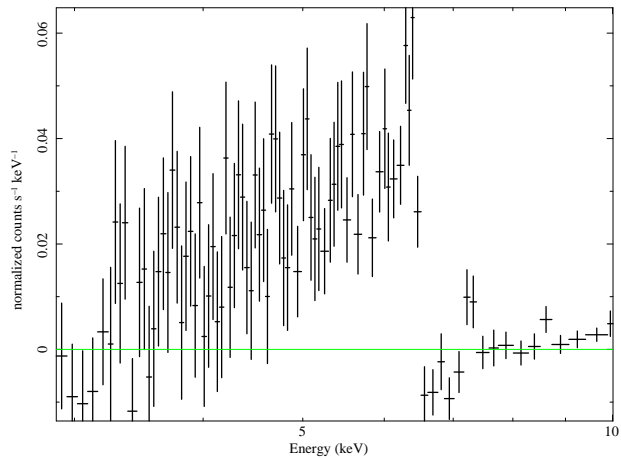}
\caption{Residuals to a model B consisting of a power law and a Gaussian line fitted in the 3-10~keV band.}\label{residuals_laor}
\end{figure}

Strong  and  broad  positive  residuals  are visible  in  the  5-8~keV
band. As a first attempt, we  fitted them with a single Gaussian line.
Considering only  the hard band,  i.e.  3-10~keV, the $\chi^2$  of the
fit  after  the  inclusion  of  the  line is  still  very  high,  i.e.
\chiqua=1.93 (model  B in Table~\ref{tab_mod1}).  The  line energy was
measured to  be at $6.15^{+0.15}_{-0.13}$~keV and  the resulting width
of  the line  of  was 0.26$^{+0.08}_{-0.04}$~keV,  with an  equivalent
width  200$^{+60}_{-20}$~eV.   Figure~\ref{residuals_laor}  shows  the
shape of the residuals to this model between $\sim$5 and 7~keV.  These
results indicate  that the spectral  feature has a broad  and slightly
skewed profile. In order to  further study the broad emission line, we
used the {\it  KYGline} model for an iron  emission line produced from
an  irradiated disc  around a  Black Hole  developed by  Dov{\v c}iak,
Karas \& Yaqoob  (2004). This model is parametrised  by the energy of
the line, the spin of the Black Hole {\it a/M}, the inclination of the
line of sight with the polar axis of accretion disc, $i$, and the size
of the emission  area on the disc, parametrised  through the inner and
outer  radii for  non--zero emissivity  of  the disc  relative to  the
horizon   radius,   $r_h(a)=1+(1-a^2)^{1/2}$.    Regarding  to   other
parameters  of  the model,  we  assumed that  ({\it  i})  there is  no
emission below the last marginal stable orbit, $r_{ms}$, and the outer
radius of the  emission ring is 400GM/c$^2$; that  ({\it ii}) the line
is emitted  from the inner  region of the  disc with the index  of the
radial dependence  of the  emissivity to 3;  and that ({\it  iii}) the
disc is emitting isotropically.  We  also fixed the energy of the line
to 6.4~keV, the  spin to its maximum value,  i.e. a/M=0.998, the outer
radius  to  400R$_g$.  We  measured  an  inclination  of the  disc  of
$24^{+4}_{-3}$ deg and an inner  radius for the emission from the disc
r$_ {in}$-r$_h<2$.  A narrow, neutral emission line with an equivalent
width  of $37\pm24$~eV  is also  included in  the fit.   Both emission
lines are statistically significant with levels higher than 99.99\%


\begin{figure*}
\includegraphics[width=50mm,angle=-90]{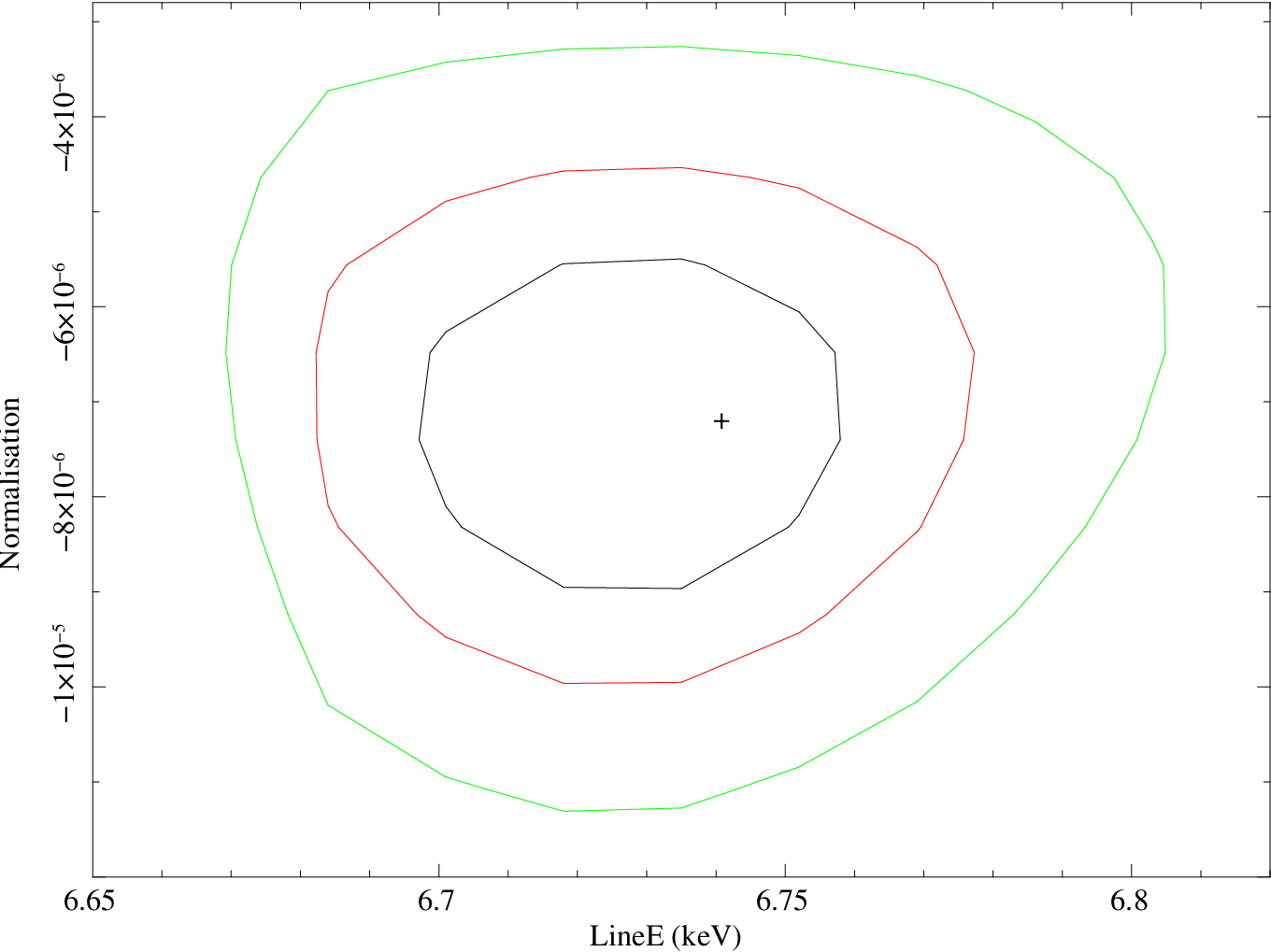}\includegraphics[width=50mm,angle=-90]{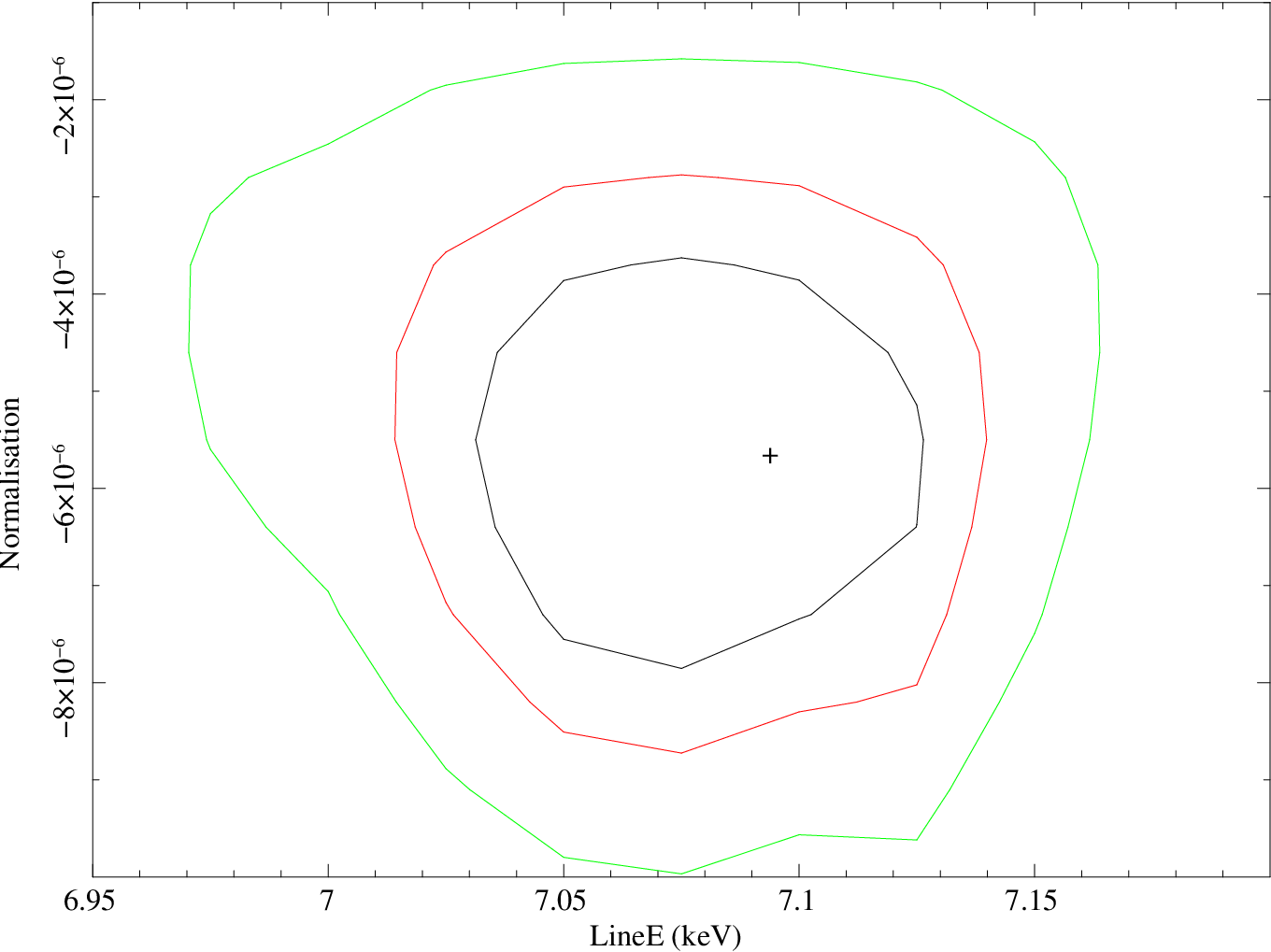}
\caption{Line energy versus  intensity contour plots for both high energy absorption lines located at $\sim6.74$~keV and $\sim7.1$~keV, left and right panel, respectively.}\label{contour_lines}
\end{figure*}

However,  negative residuals  are  still present  in the  6.5--7.5~keV
band.   These residuals  were fitted  with two  absorption  lines with
energies of $\sim$6.74~keV and  $\sim$7.1~keV.  Both lines resulted to
be  significant  according  to   the  F-test  at  $\ge$  99.99\% 
   $\Delta\chi^2=50$    and     99.85\%    $\Delta\chi^2=29$    levels,
  respectively. Contour  plot of the line energy  versus its intensity
  for each line are shown in Figure~\ref{contour_lines}.

The equivalent  width of the lines are  in both cases of  the order of
50~eV.   The  energies  of  these  two lines  suggest  that  they  are
consistent   with  the   {Fe\,\textsc{xxv}}   and  {Fe\,\textsc{xxvi}}
K$\alpha$ lines.  If this is the case, the lines would be blue-shifted
with velocities $2000^{+700}_{-1700}$ and $3000^{+1800}_{-1200}$ km/s,
respectively.  These  lines should have  been originated in  a largely
ionised material. We  therefore used the PHASE model  to determine the
properties  of the  absorbing  gas (model  C in  Tables~\ref{tab_mod1}
and~\ref{tab_mod2}). The  absorber is  required at a  confidence level
$>$99.99\%  $\Delta\chi^2=5$, according  to the F-test.  The warm
absorber model properly reproduces the absorption features observed in
the 6.5--7.5~keV and other weaker lines at higher energies, related to
other energy  levels of Fe\,\textsc{xxvi}, that  are not statistically
significant in  a line  by line analysis.   We measured  an equivalent
Hydrogen   column   for   the    ionised   gas   of   the   order   of
10$^{23}$~cm$^{-2}$     and     an     ionisation     parameter     of
$\log{U}=3.48^{+0.08}_{-0.11}$.  The velocity of the outflowing gas is
compatible with  the measurements obtained for  the absorbing Gaussian
lines, v=$1100^{+1800}_{-400}$~km/s.  The  large uncertainties are the
result of the limited spectral resolution of our data.

\subsubsection{The whole energy band}

\begin{figure}
\includegraphics[width=60mm,angle=-90]{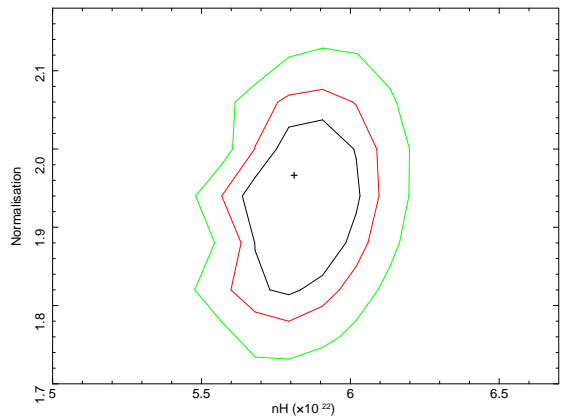}

\caption{ Equivalent  Hydrogen column
  density  of  the  cold  absorption component  versus  the  conferring
  factor contour plot}\label{cont_norm}
\end{figure}

\begin{figure}
\includegraphics[width=90mm,angle=0]{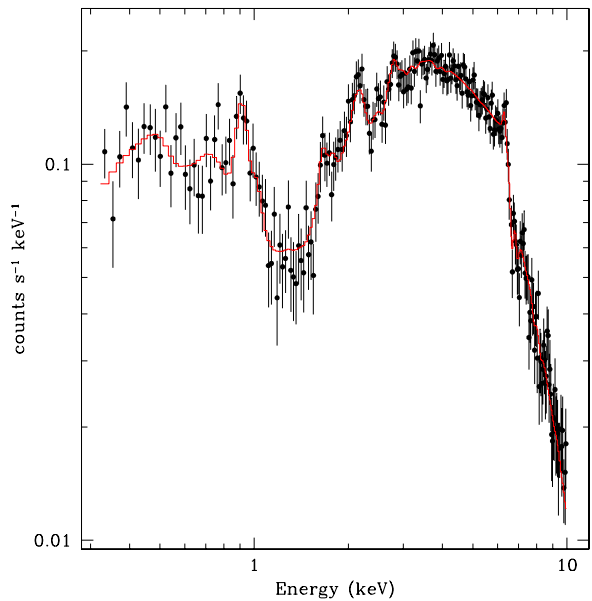}
\caption{ Observed ESO~323-G077  \pn\ spectrum and  the best fit model.}\label{wholemodel}
\end{figure}

Finally,  we modelled  the  spectrum  in the  whole  energy band.  The
presence of the  broad iron emission line also  implies a contribution
of  the reflected power  law by  a Compton  reflection from  a neutral
material of  the disc.  To model  this continuum emission  we used the
{\it  pexrav}  model in  Xspec  (Magdziarz  \&  Zdziarski 1995).   The
reflection component was relativistically convolved with the intrinsic
emissivity from the disc using the {\it KYConv} model.  Therefore, the
final best-fit model includes a  neutrally absorbed power law plus the
reflection component (R=$<0.7$) to  account for the continuum emission
of the hard energy band; another power law which accounts for the soft
band  continuum  emission,  two  different warm  absorbers,  the  four
emission  lines present in  the soft  energy band  and the  narrow and
broad components  of the iron line.  The value of  the free parameters
and   goodness  of   the  fit   are  shown   in  Tables~\ref{tab_mod1}
and~\ref{tab_mod2},  labeled  as  model  D.   Additionally,  the
  normalisation of the power law associated to the soft energy band is
  $9.8^{+0.4}_{-0.5}\times10^{-5}$ photons keV$^{-1}$cm$^{-2}$s$^{-1}$
  at 1  keV, around  2\% of the  normalisation value of  the hard-band
  power      law,      $4.95^{+0.09}_{-0.07}\times10^{-3}$     photons
  keV$^{-1}$cm$^{-2}$s$^{-1}$ at  1 keV.  Figure~\ref{cont_norm} shows
  the significance  contour levels  of the equivalent  Hydrogen column
  density  of  the  cold  absorption  component  versus  the  covering
  factor.  The  best-fit model superimposed to  the observed spectrum
is  shown in  Figure~\ref{wholemodel}.  We  have also  checked  if the
presence of  the relativistic  iron line is  still robust  within this
model by removing the {\it KYGline} component: the goodness of the fit
worsens significantly, with a value of the $\chi^2=624$ for 430 dof.

\begin{table*}
  \caption{Values of the parameters and goodness of the fits of the
  different models applied to the \epic\ spectra of ESO~322-G077. WA indicates warm absorption component.
  }\label{tab_mod1}

\begin{minipage}{280mm}
 \begin{tabular}{@{}lllcclllllll@{}}
  \hline
   Model   & PowerLaw   &   Cold Absorption  &  \multicolumn{1}{c}{ WA}  &  \multicolumn{1}{c}{ WA}  &  \multicolumn{5}{c}{ Broad Iron Line}  & \multicolumn{2}{c}{Goodness} \\ 
           & $\Gamma$      &    n$_{H}$ & {\it Low} & {\it High} &  E & $\sigma$   &   $i$/EW     & rin-rh         & a/M    & $\chi$  & DOF      \\
& &  10$^{22}$cm$^{-2}$ & {\it ionisation}& {\it ionisation} & keV & keV &  {\it deg}/eV \\
\hline

A & 1.90$^{+0.07}_{-0.06}$ & $6.34^{+0.18}_{-0.17}$ & $\surd$  & & & & & &  & 413 & 285 \\
B & 2.33$^{+0.10}_{-0.09}$ & $8.7^{+1.2}_{-1.0}$    &  & & 6.15$^{+0.15}_{-0.13}$ & 0.26$^{+0.08}_{-0.04}$ & $200^{+60}_{-20}$ & & & 339 & 216 \\ 
C & 1.95$^{+0.03}_{-0.04}$ & $6.28^{+0.14}_{-0.17}$ & $\surd$ & & 6.4f  & - & $24^{+4}_{-3}$ & $<2.0$ & 0.9982f & 544 & 439 \\
D & 1.99$\pm0.02$         & $5.82^{+0.12}_{-0.11}$   & $\surd$ & $\surd$ & 6.4f & - & $26^{+2}_{-4}$ & $<1.6$ & 0.9982f & 407 & 424 \\
D$^\dagger$ &  1.99$\pm0.02$         & $5.82^{+0.10}_{-0.13}$   & $\surd$ & $\surd$ & 6.4f & - & $26^{+3}_{-4}$ & $<1.3$ & $>0.86$ & 406 & 423  \\
 \hline \\
\end{tabular}
\end{minipage}
\end{table*}

\begin{table*}
  \caption{Values of the parameters of the warm absorption components and
    goodness  of the  fits  of  the different  models  applied to  the
    \epic\ spectra of ESO~322-G077.}\label{tab_mod2}

\begin{minipage}{280mm}
 \begin{tabular}{@{}lllllllllll@{}}
  \hline
   Model        &  \multicolumn{4}{c}{ Low Ionised Warm Absorption}  &  \multicolumn{4}{c}{ High Ionised Warm Absorption}   & \multicolumn{2}{c}{Goodness} \\ 
  &        n$_{H}$ (10$^{22}$cm$^{-2}$) & $\log{U}$ &  v$_{turb}$(km/s) & v(km/s) &       n$_{H}$ (10$^{22}$cm$^{-2}$) & $\log{U}$ &  v$_{turb}$(km/s) & v(km/s) & $\chi$  & DOF      \\
\hline

A & $13^{+18}_{-8}$ & $3.16^{+0.19}_{-0.12}$ & 300f & $<3600$ & & & & & 413 &
285 \\ 
C & $9^{+5}_{-3}$ & $3.48^{+0.08}_{-0.11}$  & 300f & $1100^{+1800}_{-400}$& & & & &
544 & 439 \\ 
D & $1.9^{+0.4}_{-0.3}$ & $2.14^{+0.06}_{-0.07}$ & $<140$
&  3200$^{+600}_{-200}$ &  $13^{+8}_{-4}$  & $3.26^{+0.19}_{-0.15}$  &
$500\pm400$ & 1700$^{+600}_{-400}$ & 407 & 424 \\
D$^\dagger$ &  $1.9^{+0.4}_{-0.3}$ & $2.14\pm0.06$ & $<140$
&  3200$^{+600}_{-300}$ &  $13^{+8}_{-4}$  & $3.23^{+0.18}_{-0.14}$  &
$500^{+400}_{-300}$ & 1600$^{+600}_{-300}$ & 406 & 423 \\
\hline \\

\end{tabular}
\end{minipage}
\end{table*}

\begin{table}
 \centering
  \caption{Soft    energy    band   emission    lines    energy, equivalent widths,
fluxes, and significance according to the F-test.}\label{tab_emlines}
  \begin{tabular}{@{}lllll@{}}
  \hline

Energy & EW  & Flux   & F-test & $\Delta\chi^2$ \\ 
keV & eV & $10^{-15}$\cgs \\ \hline
0.93 & $190^{+120}_{-20}$ & $18^{+4}_{-2}$ & $>99.99\%$ & 190\\
1.04 & $59^{+15}_{-16}$ & $3^{+3}_{-2}$ &  $>99.99\%$ & 25\\
1.68 & $36^{+13}_{-14}$ & $<6$ & $95.9\%$ & 6\\
2.84 & $18^{+9}_{-6}$ & $22^{+11}_{-19}$ & $99.0\%$  & 9 \\
 \hline
\end{tabular}

\end{table}




\subsection{Fluxes and luminosities}

The  derived flux  and luminosity,  based on  the best  fit  model and
referred   to    the   \pn\   data,   in   the    2-10~keV   band   are
$9.9^{+0.2}_{-0.5}\times10^{-12}$~erg~cm$^{-2}$s$^{-1}$              and
$7.5^{+0.2}_{-0.4}\times10^{42}$~erg/s, respectively.  In the soft
band,   the   corresponding  flux   and   luminosity  are   0.5-2~keV,
$(2.3^{+0.5}_{-0.4})\times10^{-13}$~erg~cm$^{-2}$s$^{-1}$                            and
$(5.1^{+1.1}_{-1.0})\times10^{42}$~erg/s.

We  investigated a  possible  short--term variability  of the  source,
i.e. within  the observation. We used  the {\it clean}  \pn\ events to
derive the light-curves in different  energy bands. There is no clear
sign of variability within the observation, neither in the soft nor in
the hard band.  
Only {\it RXTE}  and {\it Integral} previously observed  the source in
the hard band.  ESO323-G077 is included in the serendipitous survey of
{\it RXTE} obtained during satellite re-orientations (Revnivtsev et al.
2004) during  the 1996-2002 period.  However, the  observation date is
not  specified  in the  catalogue.   Based  on  the flux  measurements
performed by  Sazonov \& Revnivtsev  (2004), we calculated  a 2-10~keV
flux of  $\sim9\times10^{-12}$erg~cm$^{-2}$s$^{-1}$.  Similar result is
obtained when  {\it Integral} luminosity is  considered. Therefore, no
short or long-term variability can be deduced from the present data.


\section{Discussion}

\subsection{ The origin of the soft excess}

Figure~\ref{residuals_fe} shows the conspicuous {\it soft excess}
  of \eso.  The  origin of this feature in Type-I  objects is still an
  open issue.  In  general, the shape of the {\it  soft excess} can be
  fitted  by  a thermal  component,  i.e.  black-body,  bremsstrahlung
  emission. However,  the temperatures  measured for large  variety of
  AGN (i.e.  with different intrinsic luminosities, Black Hole masses,
  accretion rates...)  are of  the same order of magnitude (Piconcelli
  et al. 2005).   This result is difficult to  explain in any scenario
  related to a continuum emission  from an accretion disc.  A probably
  more  realistic scenario explains  the soft  excess emission  due to
  absorption or  emission processes, associated  to atomic transitions
  seen  blurred to  relativistic  effects (Gierli{\'n}ski  \& Done,  2004,
  Crummy et al.   2006; Schurch \& Done, 2006).   Although, the origin
  of the {\it  soft excess} of \eso\  is out of the scope  of the paper
  some different  scenarios are discussed.  In  particular, fits using
  thermal  models  (i.e.   {\it  blackboby,  bremsstrahlung},...)   to
  resemble  the  soft  excess   statistically  failed.   A  power  law
  accounting for the  soft excess provided good results,  as showed in
  previous section,  i.e. the partial covering  scenario.  The typical
  scenarios for  Seyfert 1 galaxies  described above is not plausible
  due to the  presence of the cold absorber,  which according to these
  modelisations  should  be placed  close  to  the  {\it soft  excess}
  emission source, i.e.  the  accretion disc, yielding therefore a very
  unlikely scenario. Within the  partial covering approach, we checked
  the  robustness  of the  hypothesis  by  leaving  free to  vary  the
  soft band  power law   index.   We   find  a  very   good  agreement,
  $\Gamma_s=2.03^{+0.13}_{-0.12}$    versus    $\Gamma_s=1.99\pm0.02$,
  between  both  indexes.  Moreover,  the  strength  of the  secondary
  component  is  only  2\% of  the  primary:  a  low fraction  of  the
  normalisation expected for this  scenario. If instead, the power law
  emission  is  placed  outside  the absorbing  components  (both  the
  neutral   and   the   ionised   ones),  the   fit   is   statistical
  indistinguishable  to our  best-fit model.   The resulting  index is
  $2.29\pm0.09$, in  agreement with  typical values of  Type-I objects
  (Piconcelli   et  al.    2005).    However,  this   model  is   only
  phenomenological, as it does  not correspond to any obvious physical
  scenario.

 Alternatively, the {\it soft excess}  observed in \eso\ could also be
 related to the emission/reflection  of an ionised plasma, as observed
 in many Seyfert~2 objects.  This scenario has been studied using high
 resolution  X-ray  spectra  for  individual objects  (e.g.   NGC1068,
 Kinkhabwala  et  al. 2002;  Circinus,  Sambruna  et  al. 2001;  Mrk3,
 Bianchi  et  al.  2005)  and   for  sample  of  objects  (Bianchi  et
 al. 2006). Bianchi  et al. (2006) find that the  {\it soft excess} is
 nearly ubiquitous in Seyfert~2.  Based on their high resolution X-ray
 spectral analysis,  the authors claim  that the main  contribution to
 the  {\it   soft  excess}   is  photoionised  emission   of  extended
 circumnuclear gas, illuminated by the AGN itself. The soft spectra of
 Seyfert~2 are dominated by  emission lines of highly ionised elements
 and  narrow radiative recombination  continua features  associated to
 plasma emission  of temperatures  of few eV.   In their  sample, only
 three out of  ten of the studied sources  present luminosities higher
 than  $10^{41}$~erg/s.  In  \eso,  the {\it  soft excess}  associated
 luminosity measured  is of the order  of $1.5\times10^{41}$~erg/s. If
 its {\it soft  excess} is caused by the  ionised plasma, the emission
 lines and the  radiative continua would have been  visible in the RGS
 spectra.  Therefore,  although this scenario is  somehow plausible on
 spectral modeling basis, the  large luminosity measured in our source
 makes this  origin unlikely. Finally,  an alternative origin  for the
 {\it soft excess} is  the emission from collisionally-ionised plasma.
 Interestingly, for the largely studied Seyfert~2 galaxy NGC~1068, the
 contribution of such an emission  is constrained to be less than 10\%
 (Brinkman et al.   2002).  Additionally, and as in  the hypothesis of
 the photoionised plasma as the  explanation of the {\it soft excess},
 the  lack of  detection  of  emission lines  in  the high  resolution
 spectrograph allow us to  reject this possibility.  A contribution of
 possible circumnuclear star-forming regions on \eso, can be ruled out
 due also to the large luminosity measured.

In summary,  in the case  of \eso, the  lack of high  resolution X-ray
data prevents  us from deriving any  firm conclusion on  the origin of
the  {\it soft excess}  of our  source, leaving  as the  most probable
origin  the  partial covering  scenario  with  only  a possible  minor
contribution of emission/reflection of ionised plasma.

\subsection{The ionised absorbers and emitter}

Both in  the soft  and the hard X-ray spectrum  of ESO~323-G077,
several absorption features are present.  In Section~\ref{sect:data},
we reported  the presence of  two different warm absorbers which partially
explain the noticeable absorption features in the spectrum.

\vspace{0.4cm}
{\noindent \bf The low ionised absorber}

\begin{figure}
\includegraphics[width=50mm,angle=-90]{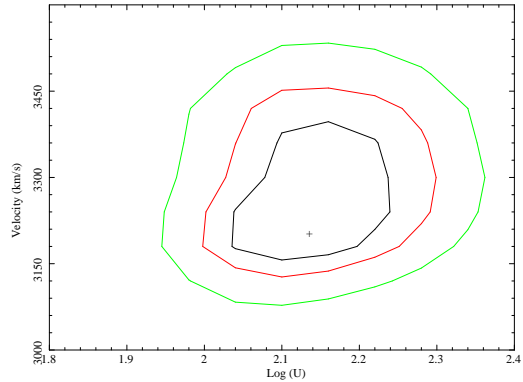}
\caption{Contour plots of the velocity  of the outflowing gas  versus  ionisation  parameter  of   the  low  ionised
  absorber.}\label{cont_wa}
\end{figure}

\begin{figure}
\includegraphics[width=80mm,angle=0]{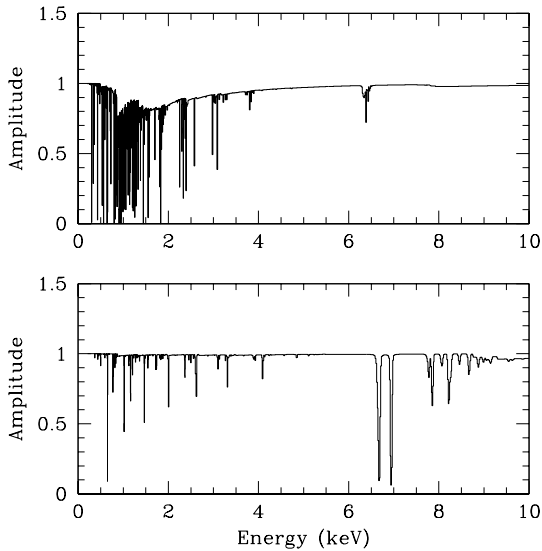}
\caption{Modelisation of the warm absorption  features of the low (upper
  panel) and high (lower panel) ionised absorbers.}\label{fig:abs}
\end{figure}

Warm absorbers are quite common  in Seyfert~1 galaxies. Around 50\% of
these  systems  present   absorbing  features  originated  by  ionised
material.   The  less  ionised  absorber  detected  in  \eso\  has  an
ionisation  parameter  of  $2.14^{+0.06}_{-0.07}$ with  an  associated
temperature  of $6.9\times10^5$~K,  an equivalent  Hydrogen  column of
$1.9^{+0.4}_{-0.3}\times10^{22}$~cm$^{-2}$  and  a turbulent  velocity
$<140$~km/s.  The  spectral analysis does  not allow to  constrain the
turbulent velocity  as no absorption  lines were resolved in  the {\it
  EPIC} spectrum. The  fit reveals that the absorption  takes place in
an outflowing  gas with  velocities between 2800  and 3400  km/s. 
  Fig.~\ref{cont_wa}  shows the contour  plot of  the velocity  of the
  outflowing gas  versus the ionisation  parameter.The lack  of high
resolution  spectra with enough  signal--to--noise ratio,  prevents us
from determining more accurately  the outflowing gas velocity.  In the
upper panel  of Fig.~\ref{fig:abs}, the imprints of  this absorber are
shown.  The absorption  is mainly affecting the low  energy band, i.e.
$<2$~keV, but  it is  also affecting the  Fe energy  band. 

\vspace{0.4cm}
{\noindent \bf The high ionised absorber}

Recently, the  high spectral resolution  and sensitivity of  \xmm\ and
\chandra\ allow  the study  of highly ionised  absorbers in  AGN which
imprint on the high X-ray energy spectra. Even if less common than the
moderately  and low  ionised absorbers,  these more  extreme absorbers
have been  reported for several  objects (e.g.  PG1211+143,  Pounds \&
Reeves, 2007, Reeves  et al. 2008; PG2112+059, Schartel  et al.  2005,
2007;  NGC1365,  Risaliti et  al.   2005;  PG1402+261,  Reeves et  al.
2004;  PG1001+054, Schartel et  al 2005;  PG1535+547, Schartel  et al
2005 ) \eso\ unambiguously hosts a high ionised absorber.  In the hard
band,  two absorption  lines, associated  with  {Fe\,\textsc{xxv}} and
{Fe\,\textsc{xxvi}}, are  observed at  a confidence level  higher than
99.9\%.  The  measured energies  of the lines  indicate that  they are
blue-shifted   with   associated    velocities   of   the   order   of
2000-3000~km/s.  We have also  reproduced them using the warm absorber
code   {\it  phase}.    The  ionisation   parameter  resulted   to  be
$3.26^{+0.19}_{-0.15}$    with    an    associated   temperature    of
$2.7\times10^6$~K   and  the   measured   equivalent  column   density
$13^{+8}_{-4}\times10^{22}$~cm$^{-2}$.   A global  fit  to both  lines
allowed  a more  accurate  measurement  of the  velocity  of the  gas,
v$=1700^{+600}_{-400}$~km/s.   The  parameters  of the  warm  absorber
could have been  accurately determined because of the  presence of the
two absorption lines which have been modelled jointly. The lower panel
of  Fig.~\ref{fig:abs} shows  this  absorber.  The  properties of  the
highly ionised absorbing gas are  therefore not as extreme as the ones
reported for  other AGN, (e.g.  PG0844+349): the  measured velocity of
the gas does not required  to invoke relativistic effects as its value
is  similar  to  those  of  typical warm  absorbers.   The  equivalent
Hydrogen  absorbing column  measured is  also in  the range  of common
values and finally no evidence of rapid variability has been detected.
However, the  accurate determination of  the properties of  the highly
ionised absorbing gas through the  {\it Phase} model allows us to more
firmly establish the presence  of the broad relativistic iron emission
line  in  ESO323-G077.  Although  high  energy (i.e.   $\sim7-10$~keV)
absorption lines have  been detected in a dozen  of objects (see Cappi
et al. 2006 for a recent  review), it is not common to observe several
of these  lines for the same  object.  Moreover, for  several of these
objects in which  more than one of these  absorption line are detected
in their spectra  ( PDS 456, Reeves et  al.  2003; PG1115+080, Chartas
et al.   2003; APM08279+52551, Chartas  et al.  2002, Hasinger  et al.
2002),  the  analysis  revealed  that  they  are  associated  to  warm
absorbers  with  different   properties  (i.e.   velocity,  ionisation
parameter...).  Therefore,  the case of  \eso\ is very  interesting as
the presence  of its highly ionised  absorber is more robust  as it is
probed by more than one line.   This finding for \eso\ is very similar
to the  case of MCG-6-30-15,  for which an outflow  with $\log\xi$=3.6
and v$\sim$2000~km/s is generating  two absorption lines associated to
K$\alpha$ lines  of {Fe\,\textsc{xxv}} and  {Fe\,\textsc{xxvi}} (Young
et  al.  2005).  The  best example  is, however,  the case  of NGC1365
(Risaliti et al. 2005), where four lines associated with K$\alpha$ and
K$\beta$  states of  both  {Fe\,\textsc{xxv}} and  {Fe\,\textsc{xxvi}}
were  detected  (see  also,  Netzer  et  al.   2003  for  NGC3783  and
Steenbrugge et al 2005 for NGC~5548).

\vspace{0.4cm}
{\noindent \bf The ionised emitter}

Four emission  lines have been detected  in the soft  energy band with
confidence levels  $>95\%$, see Figure~\ref{line_res}  .  The centroid
energy of the  lines have not been well determined  due to the limited
statistics, and  therefore, no uncertainties in the  energy line could
have been given.  However, based  only on the best-fit values, we have
associated each  detected line  to the most  likely specie.   The most
prominent  line  is  located  at $\sim0.93$~keV  (13.3\AA).   We  have
identified it  with the {Ne\,\textsc{ix}}  triplet, as it is  the most
prominent  one in  this spectral  region.  If  this is  the  case, the
observed  line  would  have   been  blueshifted  with  a  velocity  of
$\sim$2,000-4,000~km/s,  depending on  which  line of  the triplet  we
considered.         Another        intense       emission        line,
{Fe\,\textsc{xix}$\lambda$13.52\AA}, may  also be contributing  to the
observed emission and would also be originated in the same media.  The
second more significant line, located at $1.04$~keV (11.9\AA), is very
likely    originated   by    {Ne\,\textsc{x}$\lambda$12.13\AA}.    The
outflowing gas which is originating  this line would be the same which
is  responsible for  the line  at $\sim0.93$~keV,  as  the blueshifted
velocity is  also of  the order of  2,000~km/s.  The line  detected at
1.68~keV(7.38\AA)  is  also compatible  with  being  originated in  an
outflowing  gas  of  $\sim$2,000~km/s  and would  be  associated  with
{Fe\,\textsc{xxiii}$\lambda$7.48\AA}.  However, a more intense line is
located at  7.85\AA (Mg\,\textsc{xi}), but the  outflowing velocity of
the  gas would  be of  the order  of 20,000~km/s.   Finally,  the line
detected    at    2.84~keV(4.37\AA)    can    be    associated    with
{S\,\textsc{xvi}$\lambda$4.73\AA},  which  is  the  only  close  known
emission line.  However, in this  case, the outflowing velocity of the
gas would  be $\sim20,000$~km/s.  Alternatively, Kaspi  et al.  (2002)
detected in  the high resolution  spectrum of NGC3783  an unidentified
emission line at 4.35\AA. In our case, this line could be orginated in
the  same  outflowing material  with  a  velocity  of 2,000~km/s.   In
summary, the most  prominent lines detected in the  \eso\ spectrum are
likely associated with {Ne\,\textsc{ix}} (with a possible contribution
of {Fe\,\textsc{xix})  and {Ne\,\textsc{x}}. The  emitting media would
  be outflowing  with a velocity  of $\sim$2,000~km/s.  This  value is
  compatible with  the velocity  measured for the  outflowing material
  which is  causing the highly  ionised absorption.  Therefore,  it is
  likely  that  the  same  material  which is  absorbing  the  primary
  emission  is  also responsible  of  the  emission.   Although it  is
  possible  that  the same  gas  is  also  originating the  other  two
  emission lines, it is necessary  to force the identifications of the
  detected  lines with  rare species.   On  the other  hand, if  these
  emission  lines  are  identified   with  the  most  probable  lines,
  i.e. Mg\,\textsc{xi} and {S\,\textsc{xvi}}, then the velocity of the
  outflowing  gas  is abnormally  high,  (20,000~km/s). Moreover,  the
  abundance deduced if these  are actually the correct identifications
  would be  also abnormal, with  largely sub-solar abundances  of Iron,
  Oxygen,  Silicon   and  Nitrogen.  Therefore,  the   fact  that  the
  velocities calculated through the line identification are compatible
  one  with the  other reinforces  the  hypothesis of  the high  speed
  wind. It is, however, worth noting that the lines with higher energy
  are  less significant,  and  therefore no  firm  conclusions can  be
  reached with our data.

Besides this,  it is surprising that the  strongest emission line
  detected is  associated with  the {Ne\,\textsc{ix}} triplet,  but no
  hint of  {O\,\textsc{vii}} is detected in the  spectrum.  A possible
  explanation is that the metallicity  of the gas is largely different
  from the solar one.   Super-solar metallicities have been measured in
  Narrow Line Seyfert  1 (NLS1).  In fact, the  same features observed
  in  \eso\ (i.e.   {Ne\,\textsc{ix}}  emission line  with absence  of
  Oxygen lines) has  also been detected in Mrk~1239  (Grupe, Mathur \&
  Komossa,  2004).   A similar  example  of  super-solar abundance  was
  measured  in X-rays  for Mrk~1044  (Fields  et al.   2005; see  also
  Shemmer   \&   Netzer,    2002).    However,   although   super-solar
  metallicities  are common  among NLS1,  there is  only  one previous
  example  of  super-solar abundance  in  a  Seyfert~1 galaxy,  Mrk~279
  (Fields  et al.   2007).   Our  result is  very  interesting in  the
  framework of the enrichment of the intergalactic medium (Williams et
  al.   2005),  as   a  similar   trend  is   observed  in   the  O/Ne
  ratio.   The amount of mass ejected  for \eso\ in the media is
  not negligible. The mass outflow rate depends on the distance of the
  warm  absorber  from the  source,  the  column  density, the  radial
  velocity of  the outflow,  the angle  of the line  of sight  and the
  angle between the accretion disc and the gas flow.  We can assume an
  angle of  the line of  sight of $\sim40^\circ$  and an angle  of the
  outflow  of $\sim60^\circ$.

The  location of the  ionised gas  in \eso\  could not  be determined,
however,  an order  of magnitude  of  this quantity  can be  estimated
assuming that  it correlates  with the square  root of  the bolometric
luminosity.   Krongold  et   al.   (2007)  calculated  accurately  the
location of  the warm  absorber of NGC~4051  at 0.-1 lt-days  from the
nucleus.  For  \eso, we estimated the bolometric  luminosity using the
correlation   given   by    Elvis   et   al.    1994,   $L_{bol}^{ESO}
\sim(2-4)\times 10^{44}$~erg/s.   Taking into account  this measurement
and  scaling it  with luminosity  of NGC~4051,  the outflowing  gas in
\eso\ would  be located at  1.5-3 lt-days.  We therefore  estimate the
mass outflow ratio to  be of the order 1.2$\times10^{-2}$M$_\odot$/yr.
  This     corresponds    to     a     kinetic    energy     of
  $\sim3\times10^{40}$~erg/s,  which   compared  with  the  bolometric
  energy of the galaxy, corresponds  to a fraction of accretion energy
  of the order of $10^{-4}$.   Considering a lifetime of $10^8$ years
for the  AGN, it corresponds to a  total mass ejected in  the media of
the  order of  1.5$\times10^{6}$M$_\odot$.  The  corresponding kinetic
energy  (considering  the  outflow  velocity of  2000~km/s)  would  be
$\sim10^{55}$~erg/s.   This  value  is  large enough  to  disrupt  the
interstellar  medium,  causing   an  overabundance  (see  Krongold  et
al. 2007 for further discussion).

\subsection{The broad Iron emission line}

 The \xmm\ hard band spectrum of ESO323-G077 reveals that the emission
 is  complex.   The most  prominent  feature in  the  hard  band is  a
 relativistically broaden  emission iron  line. According to  the last
 studies  based  on  \xmm\  and  \chandra\  data,  the  occurrence  of
 relativistic  broad iron  emission lines  in AGN  is rare  (Fabian \&
 Miniutti, 2005; Jim\'enez-Bail\'on et al.  2005). Although it is also
 worth noting that recent results  based on analysis of a large sample
 of AGN with enough signal--to--noise in the hard X-ray band show that
 relativistically  broadened  Fe K$\alpha$  lines  are indeed  common.
 Guainazzi et al. 2006 (but see also Longinotti et al. 2007 and Nandra
 et al.   2007) detected broad iron  lines in 25-50\%  of the objects.
 Nandra et al.  2007 also  finds an occurrence of relativistic broaden
 Fe line in 30-45\% of a  sample of Seyfert 1 and intermediate type of
 AGN. However,  the actual  fraction is a  matter of debate.

One important issue  is the fact that a  highly ionised absorption can
mimic the  features of a relativistic  iron line, see  for example the
case of PG1402+262 (Reeves et  al.  2004), evidenced the importance 
of an  accurate fit  of the  data.  The imprints  of a  highly ionised
absorbing component in the X-ray  spectrum of \eso\ are prominent.  At
least  two absorption  lines  associated to  the  K$\alpha$ states  of
Fe\textsc{xxv} and Fe\textsc{xxvi} are  clearly visible in the spectrum.
We   found   that   a   dense,   n$_H\sim10^{23}$~cm$^{-2}$,   ionised
($\log{U}\sim3.5$)  absorber is  responsible for  the  lines.  Besides
that,  systematic studies  of  X-ray  emission of  AGN  show that  the
neutral   and  narrow   Fe   emission  line   is  nearly   ubiquitous,
(Jim\'enez-Bail\'on  et al.   2005;  Bianchi et  al.   2004).  In  our
analysis,  both the  narrow  component  of the  iron  and the  ionised
absorption have been taken into  account and the presence of the
broad  iron  line  is  robust. Figure~\ref{fig:con_sin_laor}  shows  a
comparison of the best fits  for the modelisation with and without the
presence of a broad iron  line.  The figure also shows graphically the
strength and significance of the line.

\begin{figure}
\includegraphics[width=80mm,angle=0]{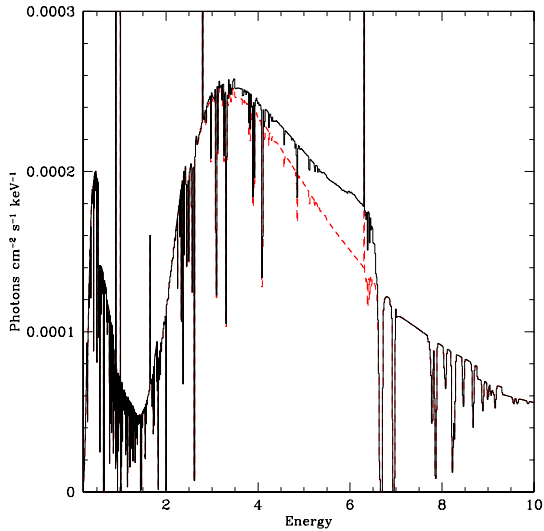}
\caption{  Best fit  model of  \eso\ in  solid line  and  fitted model
  excluding  the  relativistically broad  emission  line component  in
  dashed line. }\label{fig:con_sin_laor}
\end{figure}

The complexity of the emission in the hard band and the quality of the
present \xmm\  spectral data do not  allow us to  investigate in great
detail the parameters  of the broad emission line.   However, the
  analysis indicates that the broad emission line detected in \eso\ is
  one of the most prominent ones observed so far.  The EW of the line,
  when   a   broad  Gaussian   line   is   fitted,   resulted  to   be
  $200^{+60}_{-20}$~eV. This  value is  higher than the  mean EW  of a
  local sample of Seyfert~1 galaxies, $<EW>=91\pm13$~eV (Nandra et al.
  2007), but  comparable to the values of  relativistically broaden Fe
  lines (e.g.  EW(NGC3516)=$394^{+72}_{-63}$~eV Nandra et al.  (2007);
  EW(MGC-6-30-15)=$198^{33}_{-30}$~eV    Nandra   et    al.    (2007);
  EW(4U1344-60)=$393^{122}_{-107}$~eV   Piconcelli   et   al.    2006;
  Jim\'enez-Bail\'on et al. 2006); see other examples in Nandra et al.
  2007). When the {\it KYGline}  model is considered, the EW measuered
  resulted to be $450^{+15}_{-110}$~eV.  This value is slighly higher,
  but of  the same order of  magnitude than the EW  measuered for the
  prototypical MGC-6-30-15  using the same  model, ($\sim270\pm10$~eV,
  Dov{\v  c}iak, Karas  \& Yaqoob,  2004).  We  have  investigated the
  possibility of an  overabundance of iron to explain  the strength of
  the  measured line.   The effect  of the  iron abundance  is clearly
  imprinted  in   the  strength   of  the  Compton   reflection  hump.
  Unfortunately,  the  lack of  data  above  10~keV  prevents us  from
  investigating  in  detailed  this  parameter.   The  \xmm\  spectral
  analysis allows  us to derived  only an upper  limit of 0.7  for the
  reflection parameter  (R=$\Omega/2\pi$). Although this  value is not
  indicative as we  lack the higher energy spectrum  of the source, it
  is in good agreement with the  results for a sample of ten Seyfert~1
  galaxies performed with {\it Beppo-SAX}  (Perola et al. 2002). If we
  considered only the  five objects in the samples  which according to
  Nandra et al. (2007) host  a relativistically broaden iron line, the
  R  parameter  ranges  from  0.25  to  1.65 and  its  mean  value  is
  0.7. Similarly, the abundace of  iron is not well constrained in our
  analysis,  A$_{ Fe}=2.4^{+7.3}_{-1.9}$ times  the solar  value.  The
  fit hints, however, an overabundance of iron, $\sim2.5$ solar times.
  This also the case of  MGC-6-30-15 (Miniutti et al. 2007), where the
  abundance of iron is  $2.0^{+1.4}_{-0.6}$ solar times.  According to
  our best  fit model,  the line should  be originated in  a rotating
  black hole with an inclination  of around 25 degrees.  Nandra et al.
  (2007)  showed  that  the  mean  value of  the  inclination  of  the
  accretion  disc measuerd for  their sample  is $36\pm6$  degrees, in
  fairly  good agreement  with our  resuls.    Furthermore,  we have
calculated the  confidence of these two parameters,  together with the
innermost stable radius.  Figure~\ref{fig:contour}a shows the one, two
and three  $\sigma$ contour plot  of the inclination of  the accretion
disc and the inner radius for the stable emissivity.  We have also let
the spin  parameter free to vary,  but fixing the inner  radius of the
emitting  ring  to  be  at  the horizon  radius.   This  fit  presents
indistinguishable $\chi^2$  values with respect  to the one  found for
model  D and  compatible values  of the  interesting  parameters.  The
values  of  the  parameters and  goodness  of  the  fit are  given  in
tables~\ref{tab_mod1}    and~\ref{tab_mod2},    labeled    as    model
D$^\dagger$.   The measured  spin of  the  black hole  resulted to  be
$>0.86$.  The  confidence contours of the  spin of the  black hole and
the    inclination   of    the   accretion    disc   are    shown   in
Fig~\ref{fig:contour}b.  Finally, we have checked the stability of the
index of the  emissivity law. The fit of model  D, once this parameter
is left free,  does not improve significantly.  The  measured index of
the  emissivity law  was $3.2^{+0.3}_{-0.4}$.   The contours  plots of
this parameter  with the  inclination of the  disc have been  shown in
Fig~\ref{fig:contour_emis}.    These  fits   show  that   despite  the
degeneracy  among  the  parameters,  the relativistic  emission  lines
remains significant in all cases.

\begin{figure*}
\includegraphics[width=50mm,angle=-90]{cont_i_rin.ps}\includegraphics[width=50mm,angle=-90]{cont_spin_incl.ps}
\caption{Inclination of the disc versus  rin-rh (left panel)  and Black Hole spin (right panel)  contour plots using model D.}\label{fig:contour}
\end{figure*}

\begin{figure}
\includegraphics[width=50mm,angle=-90]{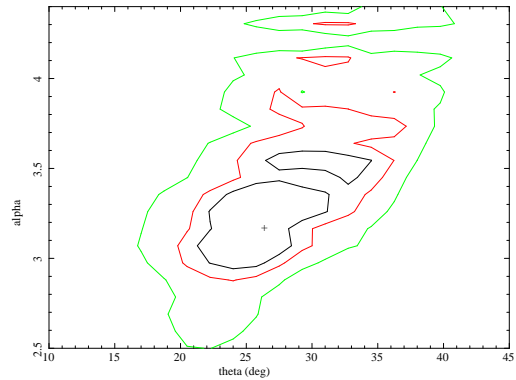}
\caption{Emissivity index versus accretion disc inclination contour plot using model D.}\label{fig:contour_emis}
\end{figure}

\subsection{Possible physical interpretation}

The \xmm\ X-ray spectrum of  \eso\ reveals a complex emission. We have
detected the  imprints of cold absorption with  an equivalent Hydrogen
column uncommonly  high for Type-I  objects.  Two warm  absorbers with
different   absorption  columns  and   ionisation  degrees   but  with
compatible outflowing velocities have  been also detected, as in other
objects (NGC985 Krongold et al.   2008; NGC~4051 Krongold et al. 2005;
NGC3783 Krongold  et al.  2003).  Several  blue-shifted emission lines
with  velocities  compatible  with  the  one  measured  for  the  warm
absorbers  are also imprinted  in the  \eso\ X-ray  spectrum.  Besides
that,  the   X-ray  analysis  evidences  the  robust   presence  of  a
relativistically broaden  iron line.  One  plausible physical scenario
which explains  the observed X-ray properties invokes  a two-phase gas
in a  conical shape outflowing along  the polar axis  (see Elvis 2000;
Krongold et al.   2007).  In our case, the outer  layer of the conical
outflow would be ionised, while the innermost layer would be much less
ionised or even neutral.  This can be obtained if a large gradient for
the  electron  density  occurs  along  the  transversal  axis  of  the
cone. The line  of sight should be such that allows  to observe in the
X-ray spectrum the imprints of the warm absorbers and the emitter with
the measured  outflowing velocity.   An inclination angle  higher than
the opening angle of the optically thick obscuring material invoked in
UM and lower than the opening angle of the cone is required to explain
both the  properties of the warm absorption/emission  and the presence
of  the broad  iron lines.   This estimation  is also  compatible with
results based  on studies in  polarised light (Schmid,  Appenzeller \&
Burch, 2003),  in which the  inclination angle calculated is  close to
45$^\circ$.    Interestingly,  the  polarised  light  is  roughly
  perpendicular to the [OIII] extended emission (Mulchaey et al. 1996)
  suggesting  that the  ionisation  cones (normally  associated to  the
  [OIII] emission) are responsible  of the polarisation. The two warm
absorbers fitted  to the spectrum suggests that  either the outflowing
absorber has only  two differentiated phases or a  continuum degree of
ionisation  but only  two components  could have  been  modeled.  This
geometry  also explains  the  emission lines  observed  with the  same
blue-shifted luminosity.  The innermost  neutral layer is evidenced by
the neutral absorption measured.  At  the same time, this component is
also blocking the redshifted  emission associated to the opposite side
of the  cone, explaining  why only the  blue-shifted component  in the
emitter is observed.

\section{Conclusions}

We reported the results on the analysis of the X-ray spectral emission
of the  Seyfert~1 galaxy \eso. The  \xmm\ data of the  source reveal a
complex spectrum with prominent  absorption and emission features. The
continuum emission  can be  explained with a  power law with  an index
roughly similar to  those of Seyfert~1 galaxies, $\Gamma=1.99\pm0.02$.
On        the        contrary,        the       cold        absorption
(n$_H=5.85^{+0.12}_{-0.11}\times10^{22}$~cm$^{-2}$)    is   large   if
compared with common values of  type I objects.  In addition, two warm
absorption components  have been also  detected in the  spectrum.  The
less ionised  absorbing gas with  $\log{U}=2.14^{0.06}_{-0.07}$ and an
equivalent               Hydrogen               column              of
$1.9^{+0.4}_{-0.3}\times10^{22}$~cm$^{-2}$   was    measured   to   be
outflowing with  a velocity in  the 2,000-3,000~km/s range.   The high
ionised       absorber,       ($\log{U}=3.26^{+0.19}_{-0.15}$      and
n$_H=13^{+8}_{-4}\times10^{22}$~cm$^{-2}$) presents a lower outflowing
velocity, v=$1700^{+600}_{-400}$~km/s. A series of four emission lines
have been  detected in  the soft energy  band. The two  most prominent
ones are  likely associated to  the {Ne\,\textsc{ix}} triplet  (with a
probable contribution of  {Fe\,\textsc{xix}}) and to {Ne\,\textsc{x}}.
If this is the case,  the associated outflowing velocities would be of
the  order  of 2,000-4,000~km/s.   Therefore,  the  material which  is
causing  the  absorption  can  also  be responsible  of  the  emission
features detected.  The fact that the strongest emission line detected
is  the  one  associated  with  {Ne\,\textsc{ix}}  while  no  hint  of
{O\,\textsc{vii}} is detected in the X-ray spectrum indicates that the
metallicity is largely different from  the solar one.  This result has
interesting   implications   to  explain   the   metallicity  of   the
interstellar medium if AGN play an important role in the feedback. The
other  two  emission lines,  less  statistically  significant, can  be
identified  with {Mg\,\textsc{xi}}  and  {S\,\textsc{xvi}}, and  hence
being  emitted   by  an  outflowing  gas  with   extreme  velocity  of
$\sim$20,000~km/s, implying  also uncommon abundance.   An outflowing
gas with a  velocity of 2,000~km/s would relate one  of the lines with
an  unidentified  transition.  However,  the  low  statistics  of  the
spectrum in this  region does not allow us  to reach firm conclusions.
Interestingly,  apart  from the  narrow  Iron  emission  line a  broad
component has been  also detected in the \eso\  spectrum. The analysis
shows that  the presence of this  line is robust and  it is compatible
with being originated  in the accretion disc of  a rotating Black Hole
with an  inclination of around $25^\circ$.   A plausible physical
  model for  this source consists  of a two-phase outflowing  gas with
  cone-like structure originated in the inner region of the AGN with a
  velocity of the order of  2,000-4,000~km/s. The inner layer would be
  less  ionised (or  even  neutral) than  the  outer one  in order  to
  reproduced  the  observed  X-ray  properties. The  system  would  be
  observed with an inclination angle  higher than the opening angle of
  the  obscuring material  invoked  in  UM and  lower  than the  angle
  subtended by the cone. This  scenario is roughly compatible with the
  results form the relativistic iron  line and with the studies of the
  sources  in  polarised  light  in  which  an  inclination  angle  of
  $45^\circ$ is addressed.  

\section*{Acknowledgements}

The authors  kindly thank  the anonymous referee  for the  useful comments  and suggestions  that significantly  improved the  paper. We also thank Matteo  Guainazzi for very useful discussions.   EJB and YK acknowledge support from the Faculty of the European Space Astronomy Centre (ESAC) and  warmly  thank  the hospitality  of  the  XMM-Newton Science  Operation Center.

\end{document}